\documentclass[preprintnumbers,amsmath,amssymb,floatfix,12pt,prd,superscriptaddress,nofootinbib]{revtex4}
\usepackage{graphicx}
\usepackage{epsfig}
\usepackage{bm}
\usepackage{amsfonts}
\usepackage{subfigure}

\def\bea{\begin{eqnarray}}
\def\eea{\end{eqnarray}}

\begin{document}
\begin{center}
\LARGE {\bf Warm-polytropic inflationary universe model}
\end{center}
\begin{center}
{M. R. Setare $^{a}$\footnote{E-mail: rezakord@ipm.ir }\hspace{1mm}
, M. J. S. Houndjo $^{b}$\footnote{E-mail: sthoundjo@yahoo.fr
}\hspace{1mm},
V. Kamali $^{a}$\footnote{E-mail: vkamali1362@gmail.com}\hspace{1.5mm} \\
 $^a$ {\small {\em  Department of Science, Campus of Bijar, University of  Kurdistan  \\
Bijar, IRAN}}\\
 $^{b}${\small {\em Institut de Math\'{e}matiques et de
Sciences Physiques (IMSP)\\01 BP 613 Porto-Novo, B\'{e}nin} }}
\\

\end{center}


\begin{center}
{\bf{Abstract}}\\
 In the present paper we study warm inflationary universe models in the context of a polytropic
 gas. We derive the characteristics of this
model in slow-roll approximation and develop our model in two cases,
1- For a constant dissipative parameter $\Gamma$. 2- $\Gamma$ as a
function of scalar field $\phi$. In these cases we will obtain exact
solution for the  scalar field and Hubble parameter. We will also
obtain explicit
 expressions for the tensor-scalar ratio $R$, scalar spectrum index $n_s$ and
its running $\alpha_s$, in slow-roll approximation.
 \end{center}
PACS Number(s): 98.80.Cq\\

\newpage

\section{Introduction}
As is well known, evidence from the cosmic microwave background
radiation indicates the early universe underwent an accelerating
phase i.e, the inflationary epoch. Moreover an inflationary phase is
one of the most compelling solution to many long-standing problems
of the standard hot big bang scenario, for example, the flatness,
the horizon, and the monopole problems, among others \cite{1,2}.
Scalar field as a source of inflation provides the causal
interpretation of the origin of the distribution of large scale
structure and observed anisotropy of cosmological microwave
background (CMB) \cite{2-i,2-i1}. In standard models for inflationary
universe, the inflation period is divided into two regimes,
slow-roll and reheating epochs. In slow-roll period kinetic energy
remains small compared to the potential terms. In this period, all
interactions between scalar fields (inflatons) and  other fields are
neglected and the universe inflates. Subsequently, in reheating
period, the kinetic energy  is comparable to  the potential energy
and inflaton starts an oscillation around the minimum of the
potential losing their energy to other fields present in the theory.
So, the reheating is the end period of inflation.\\In warm
inflationary models radiation production occurs during inflationary
period and reheating is avoided \cite{4,41}. Thermal fluctuations may
be obtained during warm inflation. These fluctuations could play a
dominant role to produce initial fluctuations which are necessary
for Large-Scale Structure (LSS) formation. So the density
fluctuation arises from thermal rather than quantum fluctuation
\cite{3-i,3-i1,3-i2,3-i3}. Warm inflationary period ends when the universe stops
inflating. After this period the universe enters in radiation phase
smoothly \cite{1,2}. Finally, remaining inflatons or dominant
radiation fields created the matter components of the
universe.\\
On the other hand, the polytropic gas has been proposed as an
alternative model for describing the accelerating of the universe
\cite{12}. Polytropic equation of state has been used in various
astrophysical situations, for example it can explain the equation of
state of degenerate white dwarfs, neutron stars and also the
equation of state of main sequence stars \cite{13}, and in the case
of Lane-Emden models \cite{14,15}. The equation of state of polytropic model may be seen
in mathematical biology \cite{nn1}, statistical physics \cite{nn2} and astronomy \cite{nn3,nn31}. This model
may be useful in cosmology \cite{nn4,nn41,nn42}. In Ref.\cite{nn4} the simple models of our universe have been studied
using a generalized equation of state $P=(\alpha+k\rho^{\frac{1}{n}})\rho$. This model have two components in equation of state: 1- A linear component
$P=\alpha\rho $. 2- Polytropic component $P=k\rho^{1+\frac{1}{n}}$. The early universe without singularity may be described by this generalized equation of
state where $\alpha=\frac{1}{3}, n=1, k=-\frac{4}{3\rho_p}$ and Planck density $\rho_p=5.16\times 10^{99}\frac{gr}{m^3}$. For a given generalized polytropic equation of state the inflationary expansion is driven by using vacuum energy \cite{nn4}. The inflation epoch happened on a time-scale of a few Planck times $t_p=5.39\times 10^{-44}$ and the universe brings to a size $a_1=1.611\times 10^{-6} m$. After this period ($t\gg t_p$) the universe inters in radiation era.
In pre-radiation era ($a\ll a_1$), it was shown that $\rho\simeq\rho_p$ where the Planck density $\rho_p$ can represent an upper bound for density \cite{nn4}. For a constant value of density we have a phase of early inflation. In the present work we initiate warm-inflation model by using the polytropic equation of state, where the inflaton fields have an interaction with the radiation field.\\
Previously chaotic inflation in
the context a phenomenological modification of gravity inspired by
the Chaplygin gas equation of state has been studied by Bertolami et
al \cite{16}. According to this model, the scalar field, which
drives inflation, is the standard inflaton field, can be extrapolate
for obtaining a successful inflation period with a Chaplygin gas
model. After that this scenario extended to the Chaplygin inspired
inflationary model in which a brane-world model is considered
\cite{17}. In the paper \cite{18}, del Campo et al have studied Warm
inflationary universe models in the context of a Chaplygin gas
equation.\\
 Based on the above statements we are motivated to
consider early universe cosmological implications of this model and
investigate if it can inspire warm inflation like the model of
\cite{18}.
\section{The Model}
The polytropic fluid has been proposed as an alternative model for
describing the accelerating of the universe where its equation of
state is of the form
\begin{equation}\label{1}
p=K\rho^{1+\frac{1}{n}}
\end{equation}
where $K$ and $n$ are constant values called in the literature
polytropic constant and polytropic index, respectively. The polytropic
constant $K$ can take the positive value for radiation and stiff fluid, the zero value for
dust and the negative value for inflationary scenario \cite{12}. From equation of state (\ref{1}) we
can find squared speed of polytropic fluid as
\begin{equation}\label{}
v_{g}^2=\frac{dp}{d\rho}=K(1+\frac{1}{n})\rho^{\frac{1}{n}}
\end{equation}
so this fluid is stable where \cite{n4,n41}
\begin{equation}\label{}
v_g^2>0\Rightarrow~~~~K<0,-1<n<0~~~~ and~~~~~K>0,n>0,n<-1
\end{equation}
The $n=-\frac{1}{2}$ case is motivated by string theory \cite{n1,n11,n12,n13,n14,n15}.
We have the conservation equation as
\begin{equation}\label{2}
\dot{\rho}+3H(\rho+p)=0,
\end{equation}
where dots mean derivatives with respect to the cosmological time.
Now, using (\ref{1}) and (\ref{2}) we can obtain
\begin{equation}\label{3}
\rho_{pt}=[-K+Ba^{\frac{3}{n}}]^{-n},
\end{equation}
where $a(t)$ is scale factor and $B$ is positive integration constant. In this article we will not
consider the above equation as a consequence  of the polytropic equation of state (\ref{1}), but we start
by studying the modified gravity, where the gravitational dynamic is given by modified Friedmann equation as
\begin{equation}\label{5}
H^2=\kappa([-K+\rho_{\phi}^{-\frac{1}{n}}]^{-n}+\rho_{\gamma}),
\end{equation}
where $H = \dot{a}/a$ is the Hubble parameter. The $\rho_{\gamma}$
is the radiation energy density. We also assume that $\kappa = 8\pi
G/3 = 8\pi/(3m_p^2)$ ($m_p$ is Planck mass) and $\rho_{\phi}=\frac{\dot{\phi}}{2}+V(\phi)$ ($V(\phi) $ is scalar potential).
This modification of energy density is understood from an extrapolation of equation (\ref{3}) as
\begin{equation}\label{6}
\rho_{pt}=[-K+\rho_m^{-\frac{1}{n}}]^{-n}\rightarrow [-K+\rho_{\phi}^{-\frac{1}{n}}]
\end{equation}
where $\rho_{m}$ is the matter energy density \cite{16}.
The dynamics of the cosmological model in the warm polytropic
inflationary scenario is given by
\begin{equation}\label{7}
\ddot{\phi}+3H\dot{\phi}+V'=-\Gamma \dot{\phi}.
\end{equation}
and
\begin{equation}\label{8}
\dot{\rho}_{\gamma}+4H\rho_{\gamma}=\Gamma \dot{\phi}^{2}.
\end{equation}
where $\Gamma$ is the dissipation coefficient, which describe the
decay of the scalar field into radiation during the inflation.
The  pressure of a scalar field is expressed
as
\begin{eqnarray}\label{9}
p_\phi=\frac{1}{2}\dot{\phi}^2-V(\phi).
\end{eqnarray}
Assuming the set of slow-roll conditions, i.e. $\dot{\phi}^2\ll
V(\phi)$ and $\ddot{\phi}\ll(3H+\Gamma)\dot{\phi}$, the Friedmann
equation (\ref{6}) reduces to
\begin{equation}\label{10}
H^2\simeq \kappa[-K+V^{\frac{-1}{n}}]^{-n},
\end{equation}
also equation (\ref{7}) reduces to
\begin{equation}\label{11}
3H(1+r)\dot{\phi}\simeq -V',
\end{equation}
where
\begin{equation}\label{12}
r=\frac{\Gamma}{3H}.
\end{equation}
By considering that during warm inflation the radiation production
is quasi-stable, i.e $\dot{\rho_{\gamma}}\ll 4H\rho_{\gamma}$ and
$\dot{\rho_{\gamma}}\ll \Gamma\dot{\phi}^{2}$, from Eq.(\ref{8})
we obtain
\begin{equation}\label{13}
\rho_{\gamma}=\frac{\Gamma \dot{\phi}^{2}}{4H}.
\end{equation}
Using the Stefan-Boltzmann formula, we have $\rho_{\gamma}=\sigma
T_{r}^{4}$, where $\sigma$ is the Stefan-Boltzmann constant, and
$T_{r}$ is the temperature of radiation. Using
Eqs.(\ref{11}),(\ref{12}),(\ref{13}), we obtain
\begin{equation}\label{14}
\rho_{\gamma}=\sigma
T_{r}^{4}=\frac{rV'^{2}}{12(1+r)^2\kappa}[-K+V^{\frac{-1}{n}}]^{n}.
\end{equation}
Introducing the dimensionless slow-roll parameters, we can
write
\begin{equation}\label{15}
\varepsilon = -\frac{\dot H}{H^2} \simeq
\frac{V'^2}{6\kappa(1+r)}V^{\frac{-1}{n}-1}[-K+V^{\frac{-1}{n}}]^{n-1}
\end{equation}
Also,
\begin{equation}\label{16}
\eta\equiv\frac{-\ddot{H}}{H\dot{H}}=\frac{[-K+V^{\frac{-1}{n}}]^{n}}{3\kappa(1+r)}[V''+\frac{V'^{2}}{V}+\frac{V'^{2}V^{\frac{-1}{n}-1}}{n(-K+V^{\frac{-1}{n}})}]
\end{equation}
Using Eqs.(\ref{11}), (\ref{14}) we find following
relation between $\rho_{\gamma}$ and $\rho_{\phi}$
\begin{equation}\label{17}
\rho_{\gamma}=\frac{r\varepsilon}{2(1+r)}\rho_{\phi}^{\frac{1}{n}+1}(-K+\rho_{\phi}^{\frac{-1}{n}})
\end{equation}
The condition under which
inflation takes place can be summarized with the parameter
$\varepsilon$ satisfying the inequality $\varepsilon<1$, which is
analogue to the requirement that $\ddot{a}>0$. This condition could
be written as
\begin{equation}\label{18}
\rho_{\phi}^{\frac{1}{n}+1}(-K+\rho_{\phi}^{\frac{-1}{n}})>\frac{2(1+r)}{r}\rho_{\gamma}
\end{equation}
Inflation ends when the universe heats up at a time when
$\varepsilon\simeq1$, which implies
\begin{equation}\label{19}
\rho_{\phi}^{\frac{1}{n}+1}(-K+\rho_{\phi}^{\frac{-1}{n}})\simeq\frac{2(1+r)}{r}\rho_{\gamma}
\end{equation}
The number of e-folds at the end of inflation is given by
\begin{equation}\label{20}
N = \int Hdt = \int\frac{H}{\dot\phi}d\phi =
-3\kappa\int_{\phi_\ast}^{\phi_f}\frac{[-K+V^{\frac{-1}{n}}]^{-n}}{V'}(1+r)d\phi
\end{equation}
\section{Perturbations}
In this section we will study the perturbations for our model. In quantum cosmology the interesting primary quantity
is the curvature perturbation spectrum which can be extract from correlation of two quantum fields in the same time. The spectrum is a measure
of size of field fluctuations. The
density of perturbation $\delta_H$  in term of quantum fluctuation $\delta\phi$ is given by
\begin{equation}\label{}
\delta_H\sim\delta N\sim H\delta t\sim H\frac{\delta t}{\delta\phi}\delta\phi\sim\frac{H}{\dot{\phi}}\delta\phi
\end{equation}
From Refs.\cite{4,n2,n21,n22} the density perturbation may be written as
\begin{equation}\label{21}
\delta_H=\frac{2H}{5\dot{\phi}}\delta \phi
\end{equation}
Using Eqs.(\ref{6}) and (\ref{7}) we can write
\begin{equation}\label{22}
\delta_{
H}^2=\frac{36(1+r)^2}{25V'^{2}}\kappa^{2}[-K+V^{\frac{-1}{n}}]^{-2n}\delta
\phi^{2}
\end{equation}
when dissipation is large, the dissipation coefficient $\Gamma$ is
much larger than the rate expansion $H$, so
$r=\frac{\Gamma}{3H}\gg1$. In warm inflationary scenario, the fluctuations of scalar field $\delta\phi$ are not generated by
quantum fluctuations but the fluctuations are generated by thermal interaction with the radiation fields, therefore as have been shown by Taylor
and Berera \cite{tb}, we have
\begin{equation}\label{23}
(\delta \phi)^{2}\simeq \frac{k_F T_r}{2\pi^{2}}
\end{equation}
where $k_F$ is wave-number and given by $k_F=H\sqrt{3r}$. Now using
Eq.(\ref{23}) we can rewrite Eq.(\ref{22}) as following
\begin{equation}\label{24}
\delta _{H}^2\simeq\frac{18\sqrt{3}}{25\pi^{2}V'^{2}}[\kappa
r(-K+V^{\frac{-1}{n}})^{-n}]^{\frac{5}{2}}=\frac{36\sqrt{3}r^{\frac{5}{2}}\kappa^2T_r}{2\pi^2V'^{2}}[-K+V^{-\frac{1}{n}}]^{-3n}
\end{equation}
We find spectral index $n_s$ (where $n_s-1=\frac{d\ln\delta_H^2}{d\ln k }$)
\begin{eqnarray}\label{25}
n_s\approx 1+2\tilde{\eta}-5\tilde{\epsilon}+\zeta
\end{eqnarray}
where
\begin{eqnarray}\label{26}
\tilde{\epsilon}=\frac{V'^2V^{-\frac{1}{n}-1}}{6\kappa r}[-K+V^{-\frac{1}{n}}]^{n-1},
\end{eqnarray}
\begin{eqnarray}\label{27}
\tilde{\eta}=\frac{[-K+V^{-\frac{1}{n}}]^n}{3\kappa r}(V''+\frac{V'^2}{V}+\frac{V'^2V^{-\frac{1}{n}-1}}{n(-K+V^{-\frac{1}{n}})})
\end{eqnarray}
and
\begin{eqnarray}\label{28}
\zeta=-\frac{2V'[-K+V^{-\frac{1}{n}}]^n}{3\kappa r}(\frac{V'}{V}+\frac{5r'}{4r}+\frac{V'V^{-\frac{1}{n}-1}}{n[-K+V^{-\frac{1}{n}}]})
\end{eqnarray}
In the above equation we have used the relation between number of
e-folds and interval in wave-number  $d\ln k=-dN$. Running of the
scalar spectral index
\begin{eqnarray}\label{29}
\alpha_s=\frac{dn_s}{d\ln k}=\frac{2V^{\frac{1}{n}+1}}{V'}[-K+V^{-\frac{1}{n}}]\tilde{\epsilon}(-5\tilde{\epsilon}'+2\tilde{\eta}'+\zeta')
\end{eqnarray}
is one of the interesting parameters which is obtained from data of
WMAP observations. From WMAP7 results,  $\alpha_s$ is approximately
$-0.038$ \cite{wmap}. Power spectrum of the tensor perturbation
during inflation epoch is given by following relation
\begin{eqnarray}\label{30}
A_g^2=6\kappa(\frac{H}{2\pi})^2\coth[\frac{k}{2T}]=\frac{3k^2}{2\pi^2}[-K+V^{-\frac{1}{n}}]^{-n}\coth[\frac{k}{2T}]
\end{eqnarray}
 where $T$ in extra factor $\coth[\frac{k}{2T}]$ denotes the temperature of the thermal background of gravitational waves \cite{ba}.
 From Ref.\cite{ba} we have , $A_g^2\propto k^{n_g}\coth(k/2T)$, so spectral index $n_g$ is obtained from power spectrum $A_g$ as
\begin{eqnarray}\label{31}
n_g=\frac{d}{d\ln k}d\ln [\frac{A_g^2}{\coth[k/2T]}]=-2\epsilon
\end{eqnarray}
Using Eqs.(\ref{24}) and  (\ref{30}), for high dissipative regime ($\Gamma>3H$), tensor-scalar ratio may be written as
\begin{eqnarray}\label{32}
R(k)=(\frac{A_g^2}{\mathcal{P}_R})|_{k=k_{*}}=\frac{V'^2[-K+V^{-\frac{1}{n}}]^{2n}}{\sqrt{3}T_r r^{5/2}}\coth[\frac{k}{2T}]|_{k=k_{*}}
\end{eqnarray}
where $\mathcal{P}_R=25\delta_{H}^2/4$. An upper bound for parameter
$R$ is obtained from observation data, ($R(k_{*}=0.002
Mp^{-1})<0.36$) \cite{k}. In next section we would like to consider
our model in two important choices for dissipative parameter
$\Gamma$. 1- $\Gamma$ is a constant parameter, 2- $\Gamma$ as a
function of scalar field $\phi$.
\section{Chaotic inflation}
In this section we will consider scalar warm-polytropic inflation
model with chaotic potential $V=\frac{m^2\phi^2}{2}$ (where $m$ is
the mass of inflaton) and in high dissipative limit.
If the inflaton $\phi$ has the polynomial interactions with the bath environment, one could take the polynomial of $\phi$ for dissipation parameter $\Gamma$ ($\Gamma=\Gamma_p\phi^p$) \cite{nm1}. The dissipation coefficient must be positive, therefor  $\Gamma_p>0$. As argued in Ref.\cite{nm1}, if the potential $V(\phi)$ is invariant under the transformation $\phi\rightarrow -\phi$, the index $p$ may be zero or even integer. $p=1,2$ cases have been considered in Refs.\cite{3-i,nm2,nm21} and general case for $p$ have been studied in Refs.\cite{18,nm1,nm3}.
We will study
this model for constant parameter $\Gamma=\Gamma_0$ and variable
dissipation parameter  $\Gamma=\alpha_p\phi^p$ .
\subsection{$\Gamma=\Gamma_0=const$}
Using Eq.(\ref{11}) and chaotic potential, we get
\begin{eqnarray}\label{33}
\dot{\phi}=-\frac{m^2}{\Gamma_0}\phi\Rightarrow~~~\phi=\phi_0\exp(-\frac{m^2t}{\Gamma_0})
\end{eqnarray}
From above equation and Eq.(\ref{10}) Hubble parameter is obtained
\begin{eqnarray}\label{34}
H=\sqrt{\kappa}[-K+(\frac{1}{2}m^2\phi_0^2 e^{-2m^2t/\Gamma_0})^{-\frac{1}{n}}]^{-\frac{n}{2}}
\end{eqnarray}
We obtain dissipation parameter $r$ from above equation and Eq.(\ref{12})
\begin{eqnarray}\label{35}
r=\frac{\Gamma_0}{3\sqrt{\kappa}}[-K+(\frac{1}{2}m^2\phi_0^2 e^{-2m^2t/\Gamma_0})^{-\frac{1}{n}}]^{\frac{n}{2}}
\end{eqnarray}
Energy density of the radiation field is related to the energy density of the inflaton as
\begin{eqnarray}\label{36}
\rho_{\gamma}=\frac{\Gamma\dot{\phi}^2}{4H}=\frac{m^2\rho_{\phi}}{2\sqrt{\kappa}\Gamma_0}[-K+\rho_{\phi}^{-\frac{1}{n}}]^{\frac{n}{2}}
\end{eqnarray}
By using Eq.(\ref{24}) scalar power spectrum  becomes
\begin{eqnarray}\label{37}
\mathcal{P}_R=\frac{\Gamma_0^2\sqrt{\Gamma_0}T_r}{4\pi^2m^2\sqrt{\kappa}V}[-K+V^{-\frac{1}{n}}]^{-\frac{7n}{4}}
\end{eqnarray}
 Tensor-scalar ratio is obtained from Eq.(\ref{32})
\begin{eqnarray}\label{38}
R(k)=\frac{18m^2V}{T_r}(\frac{\sqrt{\kappa}}{\Gamma_0})^{\frac{5}{2}}[-K+V^{-\frac{1}{n}}]^{\frac{3n}{4}}\coth[\frac{k}{2T}]
\end{eqnarray}
From WMAP7 results ($P_R\simeq 2.28\times 10^{-9}, r=0.017<0.36$) and Eqs.(\ref{37}),(\ref{38}) $V_{*}$ and $K$ become
\begin{eqnarray}\label{39}
V_{*}=2.6\times 10^{-11}\frac{T_r}{m^2}\frac{\Gamma_0^{\frac{5}{2}}}{\kappa^{29/16}\coth(\frac{k}{2T})}
\end{eqnarray}
and
\begin{eqnarray}\label{40}
K=V_{*}^{-\frac{1}{n}}-(\frac{\Gamma_0^{5/2}T_r}{4\pi^2m^2\sqrt{\kappa}V_{*}})^{\frac{4}{7n}}
\end{eqnarray}
In the above equation we have $T\simeq T_r$. If we choose $K<0$,
from above equations, we will get an upper limit for $m$
\begin{eqnarray}\label{41}
m^2<T_r\frac{\Gamma_0^{\frac{5}{2}}}{\kappa^{119/112}\coth^{3/4}(\frac{k}{2T})}2.1\times 10^{-5}
\end{eqnarray}
and a lower limit for chaotic potential
\begin{eqnarray}\label{42}
V_{*}>\frac{\kappa^{-\frac{3}{4}}}{\coth(\frac{k}{2T})}\times 10^{-6}
\end{eqnarray}
Now, using forthcoming Planck data ($R\geq 0.01$)\cite{planck,planck1}, from Eqs. (\ref{37}) and (\ref{38}) we derive
\begin{eqnarray}\label{}
V_{*}=1.8\times 10^{-13}\frac{T_r}{m^2}\frac{\Gamma_0^{\frac{5}{2}}}{\kappa^{29/16}\coth(\frac{k}{2T})}
\end{eqnarray}
When $K<0,$ we find
\begin{eqnarray}\label{}
m^2<2.7\times 10^{-10}T_r\frac{\Gamma_0^{\frac{5}{2}}}{\kappa^{119/112}\coth^{3/4}(\frac{k}{2T})}
\end{eqnarray}
We have found further constrain on the mass of Chaotic potential, from forthcoming Planck data.
In this article we have considered the stable polytropic fluid where $K<0$ and $-1<n<0$, from equation (\ref{37}) and (\ref{38}) we find
\begin{eqnarray}\label{}
-K+V^{-\frac{1}{n}}>0
\end{eqnarray}
So using the above  equation and Eq.(\ref{42}) we obtain following constraint on the value $n$
\begin{eqnarray}\label{n3}
n=\frac{1}{2q}~~~~q<0
\end{eqnarray}
Therefore, in warm-polytropic inflation model with chaotic potential, we have obtained an extra constrain (\ref{n3}) on the parameter $n$.
By using Eqs.(\ref{25}),(\ref{39}),(\ref{40}) we have  plotted parameter  $m^2$ versus the dissipation parameter $\Gamma_0$ in Fig.(1).

\subsection{$\Gamma=\Gamma(\phi)$ case}
Now we consider a situation with a power-law dissipation coefficient
\begin{eqnarray}\label{43}
\Gamma=\alpha_p\phi^p
\end{eqnarray}
where $p$ is positive and integer. By using Eqs.(\ref{11}) and (\ref{43}) scalar field is obtained
\begin{eqnarray}\label{44}
\phi(t)=[\phi_i^p-\frac{pm^2t}{\alpha_p}]^{\frac{1}{p}}~~~~~~\phi(t=0)=\phi_i
\end{eqnarray}
We find dissipative parameter $r$ in term of cosmological time
\begin{eqnarray}\label{45}
r(t)=\frac{\alpha_p}{3\sqrt{\kappa}}(\phi_i^p-\frac{pm^2t}{\alpha_p})[-K+(\frac{1}{2}m^2\phi^2)^{-\frac{1}{n}}]^{-\frac{1}{2n}}
\end{eqnarray}
In high dissipative regime ($r\gg 1$) we have
\begin{eqnarray}\label{46}
\frac{\alpha_p}{3\sqrt{\kappa}}(\phi_i^p-\frac{pm^2t}{\alpha_p})\gg(-K+(\frac{1}{2}m^2\phi^2)^{-\frac{1}{n}})^{\frac{1}{2n}}
\end{eqnarray}
From Eq.(\ref{17}) we may find a relation between energy densities  of radiation field and inflaton field.
\begin{eqnarray}\label{47}
\rho_{\gamma}=\frac{m^{p+2}}{\sqrt{\kappa}\alpha_p 2^{p/2+1}}\frac{\rho_{\phi}^{1-\frac{p}{2}}}{(-K+\rho_{\phi}^{-\frac{1}{n}})^{-\frac{1}{n}}}
\end{eqnarray}
Using Eqs.(\ref{24}) and (\ref{32}), scalar power-spectrum and
tensor-scalar ratio are given by
\begin{eqnarray}\label{48}
\mathcal{P}_R=\frac{9\sqrt{3}\kappa^{\frac{3}{4}}}{4\pi^2m^{\frac{5p+4}{4}}}\frac{T_r}{V^{\frac{4-5p}{4}}}(-K+V^{-\frac{1}{n}})^{\frac{5-12n^2}{4n}}
\end{eqnarray}
\begin{eqnarray}\label{49}
R(k)=\frac{2m^2(3\sqrt{\kappa}m^p)^{\frac{5}{2}}}{(\sqrt{3}2^p\alpha_p)^{\frac{5}{2}}}\frac{V^{\frac{4-5p}{4}}}{T_r}[-K+V^{-\frac{1}{n}}]^{\frac{-5+4n^2}{4n}}\coth[\frac{k}{2T}]|_{k=k_{*}}
\end{eqnarray}
respectively.
By using WMAP7 data ($P_R\simeq 2.28\times 10^{-9},
r=0.017<0.36$) and from Eqs.(\ref{48}),(\ref{49}) the mass of scalar
field takes the form (where $T=T_r$)
\begin{eqnarray}\label{50}
m=\frac{V^{\beta}}{a}
\end{eqnarray}
where
\begin{eqnarray}\label{51}
\beta=\frac{(4-5p)8n^2}{(5p+4)(-20n^2+5)}~~~~~~~~~~~~~~~~~~~~~~~~~~~~~~~~~~~~~~~~~~~~~~~~~~~~\\
\nonumber
a=[(\frac{5.7\times 10^{-9}}{\kappa^{3/4}T_r})^{4n^2-5}(\frac{465\kappa^{5/4}\coth(k/2T)}{(2^p\alpha_p)^{5/4}T_r})^{12n^2-5}]^{\frac{4}{(5p+4)(-20n^2+5)}}
\end{eqnarray}

When $K<0$, the upper limit for the mass is obtained from above equation
\begin{eqnarray}\label{52}
m<(\frac{(2^p\alpha_p)^{5/2}}{1.3\times 10^{30}a^{2/\beta} })^{\frac{4\beta}{(5p+4)\beta+8}}
\end{eqnarray}
This upper limit is also obtained from Planck data \cite{planck,planck1}
\begin{eqnarray}\label{}
m<(\frac{(2^p\alpha_p)^{5/2}}{1.3\times 10^{30}a'^{2/\beta} })^{\frac{4\beta}{(5p+4)\beta+8}}
\end{eqnarray}
where
\begin{eqnarray}\label{}
a=[(\frac{2.7\times 10^{-10}}{\kappa^{3/4}T_r})^{4n^2-5}(\frac{465\kappa^{5/4}\coth(k/2T)}{(2^p\alpha_p)^{5/4}T_r})^{12n^2-5}]^{\frac{4}{(5p+4)(-20n^2+5)}}
\end{eqnarray}
\begin{figure}
\includegraphics [width=60mm,height=70mm]{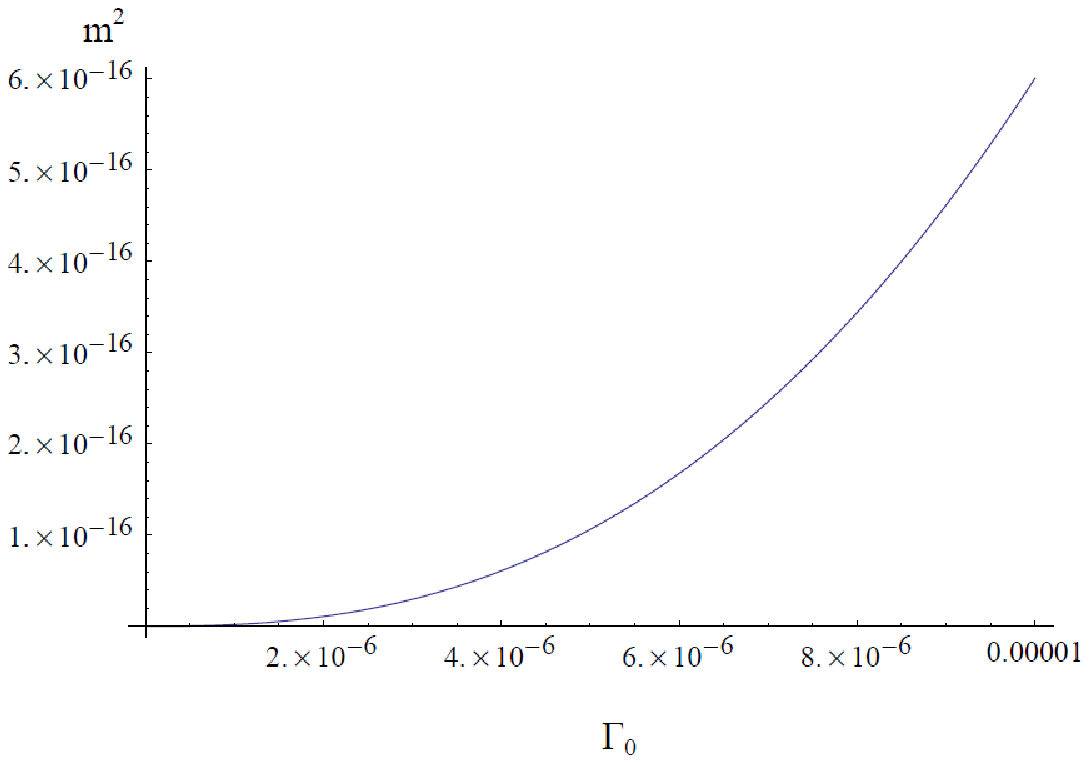
}
\caption{We plot the parameter $m^2$ in term of dissipation parameter $\Gamma_0$ where $T=T_r=2.24\times 10^{16} GeV, K_{*}=0.002 Mpc^{-1} $ and $\kappa=1$  }\label{fk3}
\end{figure}
\section{Conclusion}
In this article we have studied warm-polytropic inflationary model. We have obtained explicit
 expressions for the tensor-scalar ratio $R$, scalar spectrum index $n_s$ and
its running $\alpha_s$, in slow-roll approximation. We have developed our specific model by
 a chaotic potential, for two different cases of the dissipative parameter $\Gamma$ (where $\Gamma\gg3H$).
1- $\Gamma$ is a constant parameter and 2- $\Gamma$ is a function of scalar field.
 In these cases we have found exact solution for the  scalar field and Hubble parameter. We
also have obtained a relation between scalar field and radiation field energy densities.
 By using WMAP7 and forthcoming Planck observation data we have constrained the mass of inflaton and the chaotic potential.

\end{document}